\documentclass[vecphys]{svmult}

\usepackage{makeidx}         
\usepackage{graphicx}        
\usepackage{multicol}        
\usepackage[bottom]{footmisc}

\makeindex             

\newcommand{\be}{\begin{equation}}
\newcommand{\ee}{\end{equation}}
\newcommand{\ba}{\begin{eqnarray}}
\newcommand{\ea}{\end{eqnarray}}
\newcommand{\bc}{\begin{center}}
\newcommand{\ec}{\end{center}}

\begin{document}

\title*{Regions of star formation as $\gamma$-ray sources} \subtitle{Analytic formulae for hadronic production
in a single stellar wind 
}
\author{Diego F. Torres\inst{1}}
\institute{Instituci\'o de Recerca i Estudis Avan\c{c}ats (ICREA)
\&  Institut de Ci\`encies de l'Espai (IEEC-CSIC),
              Facultat de Ciencies, 
              Universitat Aut\`onoma de Barcelona,
              Torre C5 Parell, 2a planta, 08193 Barcelona, Spain
\texttt{dtorres@ieec.uab.es}
}
%
%
\maketitle
\index{Author1}
\index{Author2}

\begin{abstract}

We present an approximate, analytical formulae for the computation of hadronically-produced gamma-ray luminosity in the innermost regions of stellar winds. We put this into the context of other recent research on regions of star formation as gamma-ray sources.

\end{abstract}

\section{Introduction}

It has been proposed that several $\gamma$-ray sources are related
with early-type stars and their neighborhoods, since collective
effects of strong stellar winds as well as of supernova
explosions are expected to result in particle acceleration up to
energies in excess of a TeV  (see, e.g., Montmerle 1979;
Cass\'{e} \& Paul 1980; Bykov 2001,
Romero \& Torres 2003, Torres et al. 2003). Recently, several
 TeV unidentified source were detected spatially close to regions of star formation, most especially the first such source, detected  in the Cygnus
region (Aharonian et al. 2005), where a nearby EGRET source (3EG
J2033+4118) has a likely stellar origin (White \& Chen 1992; Chen
et al. 1996; Romero et al. 1999; Benaglia et al. 2001). The TeV discoveries and the increase in sensitivity at hand in the GeV and TeV regimes has rekindled the interest
in regions of star formation as gamma-ray sources.
But detecting new populations of gamma-ray sources is not a trivial task, particularly for surveying telescopes.  
For instance, a large fraction of the detections to be made by the Gamma-ray
Large Area Space Telescope (GLAST) will initially be unidentified,
and, due to their sheer number, traditional methodological approaches to identify
individuals and/or populations of $\gamma$-ray sources will
encounter procedural limitations. These limitations will hamper
our ability to classify source populations lying in the
anticipated dataset with the required degree of confidence,
particularly those for which no member has yet been convincingly
detected in the predecessor experiment EGRET. A 
new paradigm for achieving the classification of $\gamma$-ray
source populations based on the implementation of an a priori
protocol to search for theoretically-motivated candidate sources is probably needed
in order to protect the discovery potential of the sample (Torres \& Reimer 2005). Key to the any new procedure is a statistical assessment by
which the discovery of a new population can be claimed under a similar quantitative technique.
For the previously cited protocol, this is achieved by introducing that
the hypothesis to test in searching for new population of
sources is the null hypothesis against a reduced noise level.
It is useful to note that the LAT in GLAST will be in a privileged position to
actually identify new population of sources. If LAT would have an
additional order of magnitude better sensitivity, with no very
significant improvement in angular resolution, a situation similar
to the GRB case would appear, i.e., a flat distribution of
unidentified sources with a few privileged individuals only which
are extensively traced in multifrequency studies. Essentially, we
would find a $\gamma$-ray source coinciding with the position of
every member of any population under consideration. And thus, we
would lack the capability to achieve discoveries by correlation
analysis. This is, perhaps, already indicating that a next
generation high energy $\gamma$-ray mission after GLAST-LAT should
not be exclusively sensitivity-driven if no significant
improvement in angular resolution can be achieved.

\section{Hadronic processes in single and collective winds}

Regions that currently are or have been subject to a strong
process of star formation are good candidates to be $\gamma$-ray and neutrino emitters 
and may even be sites
where ultra high energy cosmic rays are produced (see Torres \& Anchordoqui 2005 for a review). Outside the
Galaxy, the more powerful sites of star formation are found within
very active galaxies such as starbursts  and Luminous or
Ultra-Luminous Infrared Galaxies, and they have also been recently analyzed 
as high energy emitters (e.g. 
Torres et al. 2003, Torres 2004, Torres et al. 2004ab, Cillis et al. 2005,  Domingo-Santamaria \& Torres 2005).
Here, we will not cover these results concerning collective effects in extragalactic scenarios, referring the reader to the mentioned publications
for details. Rather, we will make a short note on the computation of hadronic gamma-rays
in single and collective stellar winds.

Typically, a stellar wind consists of an inner region in free
expansion, followed by a larger, low dense region of hot shocked
wind (bubble), which is finally surrounded by a thin layer of
swept-up interstellar medium (ISM) (see, e.g., Lamers \& Cassinelli
1999, Ch. 12). We assume that $\gamma$-ray production is mainly
hadronic and that it occurs at the end of the free expansion phase
of the wind evolution, when the swept-up matter is comparable with
the mass in the wind driven wave. Its size is then
$R_{\rm wind}= \left({ 3 \dot{M}_\star}/{4 \pi m_p n_0 V_\infty }\right)^{1/2},
$ where $\dot{M}_\star$ is the star mass loss rate, $m_p$ is the
mass of the proton, $n_0$ is the ISM number density and $V_\infty$
is the wind terminal velocity. For a star of radius $R_\star$ the
particle density and particle velocity radial profiles of the wind
at this inner zone can be modeled by $
n(r)=\left(1-{R_0}/{r}\right)^{-\beta} {\dot {M}_\star}/
{4\pi m_p V_{\infty} r^2}, $ and $ V(r)=V_{\infty}
\left(1-{R_0}/{r}\right)^{\beta},$ respectively. The
parameter $\beta$ is $\sim 0.8 - 1$ for massive stars, $R_0$ is
given by $R_0=R_\star (1-(V_0/V_\infty)^{1/\beta})$, and the wind
velocity close to the star, $V_0$, is typically $\sim
10^{-2}V_\infty$.
As a first exercise, we can compute the resulting $\gamma$-ray luminosity of a single star with the target matter being that of the innermost region, and between
energies $E_1$ and $E_2$:
\begin{equation}
L_\gamma = \int_{R_\star}^{R_{\rm wind }}
 \int_{E_1}^{E_2} n(r) \, q_{\gamma}(E_\gamma) E_\gamma \, (4\pi r^2)dr \,
 dE_\gamma ,
\label{lum}
\end{equation}
with the differential $\gamma$-ray emissivity from $\pi^0$-decays,
$q_{\gamma}(E_\gamma)$, approximated by \be
q_{\gamma}(E_{\gamma})= 4 \pi \sigma_{pp}(E_p)
({2Z^{(\alpha)}_{p\rightarrow\pi^0}}/{\alpha})\;J_p(E_{\gamma})
\eta_{\rm A}\Theta(E_p-E_p^{\rm min}) \ee at the energies of
interest.  The parameter $\eta_{\rm A}$ takes into account the
contribution from different nuclei in the wind (for a standard
composition $\eta_{\rm A} \sim 1.5$, Dermer 1986).
$J_p(E_{\gamma})$ is the proton flux distribution evaluated at
$E=E_{\gamma}$ (units of protons per unit time, solid angle,
energy-band, and area), which is assumed as a typical power-law
$J_p(E_p)=(c/4\pi)K_p {E_p}^{-\alpha}$, with $\alpha$ being the
spectral slope and $K_p$ the spectrum normalization factor. The
cross section $\sigma_{pp}(E_p)$ for $pp$ interactions at energy
$E_p\approx 10 E_{\gamma}$ can be represented above $E_p\approx
10$ GeV by $\sigma_{pp}(E_p)\approx 30 \times [0.95 + 0.06 \log
(E_p/{\rm GeV})]$ mb (e.g., Aharonian \& Atoyan 1996).
$Z^{(\alpha)}_{p\rightarrow\pi^0}$ is the so-called
spectrum-weighted moment of the inclusive cross-section. Its value
for different spectral indices $\alpha$ is given, for instance, by
Drury et al. (1994). Finally $\Theta(E_p-E_p^{\rm min})$ is a
Heaviside function that approximately takes into account the fact
that only CRs with energies higher than $E_p^{\rm min}(r\gg
R_\star)$ will diffuse into the wind (see below).
For normalization purposes, we take the expression of the
proton energy density, which is given by \be \omega_{\rm CR} = \int
N_p(E_p) E_p dE_p = \{ 134.4 \,;\, 9.9 \, ;\, 0.8 \} K_p\; {\rm
eV}\; {\rm cm}^{-3} \equiv \varsigma \omega_{\rm CR,\odot} , \ee
where $\varsigma$ is the enhancement factor of the CR energy
density with respect to the local value, $ \omega_{\rm CR,\odot}$,
we have considered energies between 1 GeV and 20 TeV, and the
choice for the prefactor, $\{ 134.4 \,;\, 9.9 \, ;\, 0.8 \} {\rm
eV}\; {\rm cm}^{-3}$ , refers to three different slopes, 1.9, 2.0,
and 2.1, respectively. If the prefactor is referred as $f(\alpha)$
[with units of $\omega_{\rm CR,\odot}$, typically given in eV
cm$^{-3}$], the normalization is  $ K_p(\alpha) \sim {
\omega_{\rm CR,\odot} \varsigma }/{f(\alpha)} ,$ depending on the
slope of the spectrum. We assume the Earth-like spectrum to be
$J_\oplus(E)=2.2 E_{\rm GeV}^{-2.75} {\rm cm^{-2} s^{-1} sr^{-1}
GeV^{-1}} $, so that $
\omega_{\rm CR,\odot} \sim 1.5 \, {\rm eV cm^{-3}}$.
An analytic expression for the $\gamma$-ray luminosity of a single
star produced in the innermost region of its wind 
can thus be obtained if the radial dependence of the wind velocity
is ignored. This, in fact, does not introduce significant changes on
the results because a) the terminal velocity of the wind is reached
in only a few star radii,  and b) the integration extends to radii
up to $R_{\rm wind} \gg R_\star$, with the largest radii (where most of the mass is located) dominating
the sum. The expression for the approximate luminosity then reduces
to:
\begin{eqnarray}
L_\gamma \sim \frac {9.6 \times 10^{-38} \, c \, \dot{M}_\star \,
Z^{(\alpha)}_{p\rightarrow\pi^0} \, \eta_{\rm A} K_p(2) } {m_p
V_\infty \alpha} ( R_{\rm wind} - R_\star ) \; \times \; \hspace{2cm} \nonumber \\
\left[ 0.47 \ln{\frac{E_2}{E_1}} + \frac{0.06}{2 \log{e}} \left[
\left(\log \left(\frac{E_2}{\rm eV}\right) \right)^2 - \left(\log
\left(\frac{E_1}{\rm eV}\right) \right)^2 \right] \right] \; {\rm
erg\; s}^{-1},
\label{lumin1}
\end{eqnarray}
for $\alpha$=2, and
\begin{eqnarray}
\label{lumin2} L_\gamma \sim \frac {9.6 \times 10^{-38} \, c \,
\dot{M}_\star \, Z^{(\alpha)}_{p\rightarrow\pi^0} \, \eta_{\rm A}
K_p(\alpha) } {m_p
V_\infty \alpha} ( R_{\rm wind} - R_\star ) \; \times \;\hspace{4cm} \nonumber \\
\left[ \frac{0.47}{2-\alpha} \left( \left(\frac{E_2}{\rm eV}\right)^{2-\alpha} -
\left(\frac{E_1}{\rm eV}\right)^{2-\alpha} \right) - 
\frac{0.06}{\ln{10} \, (\alpha-2)} \left( \left(\frac{E_2}{\rm eV}\right)^{2-\alpha}
\ln \left(\frac{E_2}{\rm eV}\right) - \right. \right. \hspace{1cm} \nonumber \\ 
\nonumber
\left. \left.
\left(\frac{E_1}{\rm eV}\right)^{2-\alpha}
\ln \left(\frac{E_1}{\rm eV}\right) \right) - 
\frac{0.06}{\alpha^2 - 4\alpha + 4} \left( \left(\frac{E_2}{\rm eV}\right)^{2-\alpha} -
\left(\frac{E_1}{\rm eV}\right)^{2-\alpha} \right) \right] {\rm erg\, s}^{-1} \;\;\;\;\;\;\;\;\;\;\;\;\;\;\;\; \mbox{}
\end{eqnarray}
for $\alpha\ne$2, with the proton mass measured in kg, the terminal
velocity and the speed of light measured in cm s$^{-1}$, the mass
loss rate measured in kg s$^{-1}$, and the wind and stellar radius
given in cm.
Analogously, the expected $\gamma$-ray flux over an energy threshold
$E_{\rm th}$ can be evaluated taking into account the distance $D$
from the Earth to the stellar association:
\begin{equation}
F_\gamma(E_\gamma>E_{\rm th}) = \frac{1}{4\pi D^2} \int_{R_\star}^{R_{\rm
wind }} \int_{E_{\rm th}} n(r)\, q_{\gamma}(E_\gamma) \, 4\pi r^2
\, dr \, dE_\gamma \label{flux} .
\end{equation}
Also here an analytic --although approximate-- expression,
can be obtained:
\begin{eqnarray}
F_\gamma(E>E_{\rm th}) \sim  \frac{1}{4\pi D^2} \frac {6.0 \times
10^{-26} \, c \, \dot{M}_\star \, Z^{(\alpha)}_{p\rightarrow\pi^0}
\, \eta_{\rm A} K_p } {m_p V_\infty \alpha} ( R_{\rm wind} - R_\star
) \times \hspace{2cm} \nonumber \\  \left[ \frac{0.47}{\alpha-1} \left( \frac {E_{\rm th}} {\rm eV}
\right)^{1-\alpha} + \frac{0.06} {\ln
10} \left( \frac {E_{\rm th}} {\rm eV} \right) ^{1-\alpha} \left(
\frac{1}{\alpha^2 - 2\alpha + 1} +\frac 1{\alpha-1} \ln \left( \frac
{E_{\rm th}} {\rm eV} \right)  \right) \right ] \hspace{1cm}  \nonumber \\  {\rm photons \,
cm}^{-2} {\rm s}^{-1} , \hspace{3cm}  \label{ff} \nonumber
\end{eqnarray}
which is valid for all $\alpha \neq 1$. Here, $D$ must be given in
cm, whereas all other quantities must be given in the same units
as those used in Eqs. (4) to respect consistency with the
numerical factor. This formulae are useful to fix the scale: a typical single star would produce, by interaction
between cosmic rays and target matter in the innermost region of its wind, a flux one to three orders of magnitude (depending on the kind of star) below the sensitivity achieved by current telescopes in the TeV regime. Of course, this estimation does not take into account other processes, particularly those happening in binary systems where two strong stellar winds interact.


In the case of a collection of winds, a hydrodynamical model
taking into account the composition of different single stellar winds needs to be adopted. As we see below, differences with the single stellar wind case are notable. One possibility is to consider
that there are $N$
stars in close proximity, uniformly distributed within a radius
$R_c$, with a number density
$ n = {3N}/{ 4\pi R_c^3} \,. $
Each star has its own mass-loss rate ($\dot M_{i}$) and (terminal)
wind velocity ($V_i$), and average values can be defined as
$\dot M_w = {1}/{N} \times  \sum_{i}^{N} \dot M_i \,,  $  and $
V_w = \left( {\sum_{i}^{N} \dot M_i {V_i}^2}/{N\,\dot M_w}
\right)^{1/2} \,. $
All stellar winds may be assumed to mix with the surrounding ISM and
with each other, filling the intra-cluster volume with a hot,
shocked, collective stellar wind. A stationary flow in which mass
and energy provided by stellar winds escape through the outer
boundary of the cluster is established.
For an arbitrary distance $R$ from the center of the cluster, mass
and energy conservation imply 
\ba \frac{4 \pi}{3} R^3 n \dot M_w &=& 4 \pi R^2 \rho V \,, \\
\frac{4 \pi}{3} R^3 n \dot M_w \left( \frac 12 {V_w}^2 \right)&=&
4 \pi R^2 \rho V \left( \frac 12 {V}^2 + h\right) \,, \ea
where $\rho$ and $V$ are the mean density and velocity of the cluster
wind flow at position $R$ and $h$ is its specific enthalpy (sum of
internal energy plus the pressure times the volume).
%
%
Solving these equations with the corresponding boundary conditions, one obtains the density and velocity of the cluster as a function of distance from its center. This is presented in detail in the paper by Domingo-Santamaria \& Torres (2006).
Again, it is to be noted that not all cosmic rays will be able to enter
the collective wind. The difference
between an {\it inactive target}, as that provided by matter in
the ISM, and an {\it active or expanding target}, as that provided
by matter in a single or a collective stellar wind, is indeed given by 
modulation effects.
The cosmic ray penetration into the jet outflow
depends on the parameter $\epsilon \sim V R/D$,
where $V$ is velocity of wind, 
and $D$ is the diffusion coefficient. $\epsilon$ measures the ratio
between the diffusive and the convective timescale of the
particles. Taking into account such effects, 
we have studied collective wind configurations produced by a
number of massive stars, and obtained densities and expansion
velocities of the stellar wind gas that is  target for hadronic
interactions in several examples (see details in Domingo-Santamar\'ia \& Torres 2006). We have computed secondary
particle production, electrons and positrons from charged pion
decay, electrons from knock-on interactions, and solve the
appropriate diffusion-loss equation with ionization, synchrotron,
bremsstrahlung, inverse Compton and expansion losses to obtain
expected $\gamma$-ray emission from these regions, including in an
approximate way the effect of cosmic ray modulation (see Torres \& Domingo-Santamar\'ia 2005 for details on the code). Examples
where different stellar configurations can produce sources for
GLAST and the MAGIC/HESS/VERITAS telescopes in non-uniform ways,
i.e., with or without the corresponding counterparts were found.
 Cygnus OB 2 and Westerlund 1 maybe  two
associations where this scenario could be at work

\section*{Acknowledgments}

DFT has 
been supported by Ministerio de Educaci\'on y Ciencia (Spain) 
under grant AYA-2006-00530, and the Guggenheim Foundation.

\printindex

\begin{thebibliography}{99}

\bibitem{} Aharonian F. A. et al. 2005, A\&A 31, 197


\bibitem{}  Aharonian F. A. \& Atoyan, A. M. 1996, A\&A 309, 91 



\bibitem{}  Benaglia P., et al. 2001 A\&A
366, 605

\bibitem{632} Bykov, A. M. 2001, Space Sci. Rev., 99, 317

\bibitem{634} Cass\'{e}, M. \& Paul, J. A. 1980 ApJ, 237, ~236

\bibitem{641} Chen, W., White, R. L., \& Bertsch, D. 1996, A\&AS, 120, 423

\bibitem{} Cillis A. N., Torres D. F. \& Reimer O. 2005,
  ApJ {621}, 139


\bibitem{13} Domingo-Santamar\'{\i}a E. \& Torres D. F. 2005, A\&A 444, 403
\bibitem{20} Domingo-Santamar\'{\i}a E. \& Torres D. F. 2006, A\&A 448, 613

\bibitem{670}Lamers H. J. G. L. M., \& Cassinelli, J. P. 1999, Introduction
to Stellar Winds,  Cambridge University Press, Cambridge


\bibitem{680} Montmerle T. 1979, ApJ, 231, ~95


\bibitem{688} Romero G. E., \& Torres, D. F. 2003, ApJ, 586, L33

\bibitem{686} Romero, G. E., Benaglia, P., \& Torres, D. F. 1999, A\&A, 348, ~868

\bibitem{693} Torres D. F., et al. 2003, Physics Reports, 382, 303

\bibitem{} Torres D. F. 2004, ApJ 617, 966

\bibitem{} Torres D. F. \& Anchordoqui L. A. 2004, Rept. Prog. Phys. 67, 1663

\bibitem{} Torres D. F., et al. 
2004a, ApJ 607, L99

\bibitem{} Torres D. F., Domingo-Santamar\'{\i}a E., \& Romero G. E. 2004b,
ApJ 601, L75

\bibitem{} Torres D. F. \& Reimer O. 2005, ApJ 629, L41

\bibitem{} Torres D. F. \& Domingo-Santamar\'{\i}a E. 2005
Mod. Phys. Lett. A20, 2827

\bibitem{} White, R. L., \& Chen W. 1992, ApJ, 387, ~L81
\end{thebibliography}
\end{document}